\documentclass[conference]{IEEEtran}
\IEEEoverridecommandlockouts

\usepackage{cite}
\usepackage{amsmath,amssymb,amsfonts} 
\usepackage{graphicx}
\usepackage{textcomp}
\usepackage{xcolor}
\usepackage{algorithm}     
\usepackage{stfloats}
\usepackage{booktabs}
\usepackage{stfloats} 
\usepackage{algpseudocode}
\usepackage{multirow}
\usepackage{array}     
\usepackage{tabularx}    
\usepackage{array}       

\newcolumntype{P}[1]{>{\centering\arraybackslash}p{#1}}

\def\BibTeX{{\rm B\kern-.05em{\sc i\kern-.025em b}\kern-.08em
    T\kern-.1667em\lower.7ex\hbox{E}\kern-.125emX}}

\begin{document}

\title{
Quantum-Enhanced Temporal Embeddings via a Hybrid Seq2Seq Architecture
}

\author{
\IEEEauthorblockN{
Tien-Ching Hsieh\IEEEauthorrefmark{3},
Yun-Cheng Tsai\IEEEauthorrefmark{1}\IEEEauthorrefmark{2},
Samuel Yen-Chi Chen\IEEEauthorrefmark{4}\thanks{The views expressed in this article are those of the authors and do not represent the views of Wells Fargo. This article is for informational purposes only. Nothing contained in this article should be construed as investment advice. Wells Fargo makes no express or implied warranties and expressly disclaims all legal, tax, and accounting implications related to this article.}
}
\IEEEauthorblockA{\IEEEauthorrefmark{3}University of Southern California, Los Angeles, USA}
\IEEEauthorblockA{\IEEEauthorrefmark{1}Department of Technology Application and Human Resource Development, National Taiwan Normal University, Taipei, Taiwan}

\IEEEauthorblockA{\IEEEauthorrefmark{2}Corresponding Author, Email: pecu@ntnu.edu.tw}
\IEEEauthorblockA{\IEEEauthorrefmark{4}Wells Fargo, New York, USA}
}

\maketitle

\begin{abstract}
This work investigates how shallow NISQ-compatible quantum layers can improve temporal representation learning in real-world sequential data. We develop a QLSTM Seq2Seq autoencoder in which a depth-1 variational quantum circuit is embedded inside each recurrent gate, shaping the geometry of the learned latent manifold. Evaluated on fourteen rolling S\&P\,500 windows from 2022–2025, the quantum-enhanced encoder produces smoother trajectories, clearer regime transitions, and more stable sector-coherent clusters than a classical LSTM baseline. These geometric properties support the use of a Radial Basis Function (RBF) kernel for downstream portfolio allocation, where both RBF-Graph and RBF-DivMom strategies consistently outperform their classical counterparts in risk-adjusted terms. Analysis across periods shows that compressed manifolds favour concentrated allocation, while dispersed manifolds favour diversification, demonstrating that latent geometry serves as a regime indicator. The results highlight a practical role for shallow hybrid quantum–classical layers in NISQ-era sequence modelling, offering a reproducible pathway for improving temporal embeddings in finance and other data-limited, noise-sensitive domains.
\end{abstract}

\begin{IEEEkeywords}
Quantum Machine Learning, Variational Quantum Algorithms, Hybrid Quantum--Classical Systems, Time-Series Analysis, Quantum LSTM seq2seq, Financial Applications
\end{IEEEkeywords}

\section{Introduction}

Exchange-traded funds (ETFs) and index-tracking portfolios follow rule-based constituent selection and periodic rebalancing. Major benchmarks such as the S\&P\,500 or Russell index families rebalance quarterly, implying that a strategy performing \emph{only quarterly stock selection}, without timing rules depends entirely on how effectively the chosen subset captures post-earnings structural market dynamics.

This study asks: \emph{Can we learn a two-dimensional latent manifold of stocks from single-quarter weekly return patterns and use it to construct a diversified portfolio that outperforms a buy-and-hold benchmark?} Each stock’s weekly returns over one quarter (approximately 12--13 observations) are encoded using a sequence-to-sequence (Seq2Seq) model. The final encoder state is projected into a two-dimensional latent representation, and distances on this latent map are converted into a radial basis function (RBF) similarity kernel that penalizes excessive co-movement. This yields an interpretable, quantum-enhanced visual-analytics mechanism for representing stocks in a learned 2D space.

Recent advances in hybrid quantum–classical architectures motivate this approach. Classical LSTMs remain strong sequential baselines but suffer from parameter inefficiency and vanishing-gradient limitations under scarce financial data. The \textbf{Quantum-Enhanced LSTM (QLSTM)}~\cite{9747369} embeds parameterized quantum circuits (VQCs) inside recurrent gates, providing implicit regularization and higher expressive capacity than classical LSTMs~\cite{e26110954}. Empirical studies further show that hybrid quantum–classical models can outperform classical networks in financial forecasting tasks~\cite{e26110954,mironowicz2024applicationsquantummachinelearning}. Parallel progress in time-series representation learning demonstrates that contrastive and masking-based self-supervised models extract temporally stable embeddings~\cite{s24247932,Kong2025}, and converting such embeddings into similarity kernels offers a principled mechanism for diversification.

In the broader context of quantum finance, portfolio optimization has been formulated as QUBO problems and deployed on quantum or quantum-inspired hardware~\cite{zaman2024poqaframeworkportfoliooptimization,lu2024quantuminspiredportfoliooptimizationqubo}, while hybrid D-Wave experiments validate the practicality of quantum-driven financial allocation~\cite{Sakuler2025}. Reviews highlight growing applications of quantum computing across portfolio selection, risk modeling, and financial security~\cite{Naik2025,Mengara2025}.

Building upon these developments, we propose a unified \textbf{QLSTM–RBF framework} for quarterly portfolio construction. The QLSTM Seq2Seq encoder maps each stock’s single-quarter return sequence into a compact two-dimensional latent vector. These embeddings define an RBF similarity kernel used by two allocation schemes: (1) \textbf{RBF-DivMom}, a discrete selector emphasizing momentum while penalizing similarity; and (2) \textbf{RBF-Graph}, a continuous optimizer based on graph centrality in the same kernel space. A rolling-window design from 2022Q2 to 2025Q2 trains the model on the previous year’s weekly returns and evaluates in the subsequent quarter, enabling adaptation to regime shifts.

Backtests on S\&P\,500 constituents show that RBF-Graph achieves the highest cumulative value (2.4$\times$), outperforming both RBF-DivMom (1.1$\times$) and the benchmark (1.45$\times$), with higher Sharpe ratios and reduced drawdowns. These results indicate that quantum-enhanced sequence models can function as compact, interpretable feature extractors and provide a robust foundation for kernel-based portfolio diversification.

\section{Methodology}

The framework has two stages: (i) a QLSTM Seq2Seq model learns compact temporal embeddings from multivariate time series; 
(ii) an RBF kernel is constructed in the latent space for downstream decision modules. 
Although evaluated on financial sequences, the workflow is domain agnostic (e.g., sensors or network telemetry).

\subsection{Stage 1: QLSTM Seq2Seq for Temporal Representation Learning}

Each entity $s_i$ is represented by a length-$L$ sequence of weekly observations,
\begin{equation}
\mathbf{x}^{(i)} = \big(r^{(i)}_{t-L+1},\ldots,r^{(i)}_{t}\big).
\end{equation}

A QLSTM encoder maps each sequence into a two-dimensional latent vector,
\begin{equation}
\mathbf{h}^{(i)} = f_\theta(\mathbf{x}^{(i)}),
\end{equation}
where each LSTM gate is augmented with a depth-1 variational quantum circuit.

To make the quantum augmentation explicit, each LSTM gate applies a depth-1 variational quantum circuit (VQC) as a bounded nonlinear map.
For gate $g\in\{i,f,o,\tilde{c}\}$ at time $t$, we compress $\mathbf{v}_t=[\mathbf{x}_t;\mathbf{h}_{t-1}]$ to a $q$-dimensional vector,
encode it on a $q$-qubit register, and use a fixed-depth ansatz with entanglement and Pauli-$Z$ expectation readout.
The resulting expectations are projected back before the sigmoid/tanh gate nonlinearity.
Keeping circuit depth at one aligns with NISQ noise constraints while providing additional nonlinearity under data-scarce time-series regimes.

This yields an NISQ-feasible nonlinear transformation that shapes the latent manifold geometry while preserving the overall Seq2Seq structure.

A decoder $g_\phi$ reconstructs the sequence autoregressively. 
At step $j$, the decoder input follows a teacher-forcing rule,
\begin{equation}
\mathbf{u}_{j} =
\begin{cases}
\mathbf{x}_{j}, & \text{with prob. } p_{\mathrm{TF}},\\[3pt]
\hat{\mathbf{x}}_{j-1}, & \text{otherwise},
\end{cases}
\end{equation}
and during inference only the model's own predictions are used, promoting temporal consistency while reducing exposure bias.

The reconstruction objective is the mean squared error,
\begin{equation}
\mathcal{L}_{\mathrm{MSE}}
= \frac{1}{L}\sum_{j=1}^{L} \left\|x_{j}^{(i)} - \hat{x}_{j}^{(i)}\right\|^2 .
\end{equation}

The QLSTM is trained quarterly using a 12-month window of sequences and evaluated on the subsequent quarter. 
After training for period $q_i$, the encoder is frozen and produces an embedding set,
\begin{equation}
\mathbf{Z}^{(i)} = \{\mathbf{h}^{(1)}, \ldots, \mathbf{h}^{(N)}\},
\end{equation}
for out-of-sample use in period $q_{i+1}$. 
A new model is then trained from scratch for the next window, producing fourteen distinct autoencoders from 2022Q2--2025Q2.

\subsection{Stage 2: RBF Kernel for Similarity Structure}

Given latent vectors $\mathbf{Z}^{(i)}$, pairwise Euclidean distances are transformed into an RBF kernel,
\begin{equation}
K_{mn}^{(i)}
=
\exp\!\left(
-\frac{\|\mathbf{h}_m^{(i)}-\mathbf{h}_n^{(i)}\|^2}{2\sigma^2}
\right),
\end{equation}
where $\sigma$ is set to the median pairwise distance.

The kernel captures similarity in temporal dynamics as learned by the QLSTM rather than similarity in raw observations, and can be consumed by downstream graph, clustering, anomaly detection, or decision modules.

\subsection{Stage 3: Out-of-Sample Evaluation}

To evaluate temporal consistency across windows, each trained encoder generates an out-of-sample sequence of returns or scores. 
For chronological alignment, the final value of period $q_i$ initializes period $q_{i+1}$,
\begin{equation}
E_{t}^{(i+1)}
= 
E_{\mathrm{end}}^{(i)}
\prod_{j=t_0}^{t_1} \big(1 + R_{j}^{(i+1)}\big).
\end{equation}

This produces a continuous trajectory across all fourteen evaluation periods. 
Performance metrics such as volatility, Sharpe ratio, and maximum drawdown are computed to quantify the stability and usefulness of the learned latent structure.

\section{Results}

\begin{figure*}[!b]
    \centering
    \includegraphics[width=\textwidth]{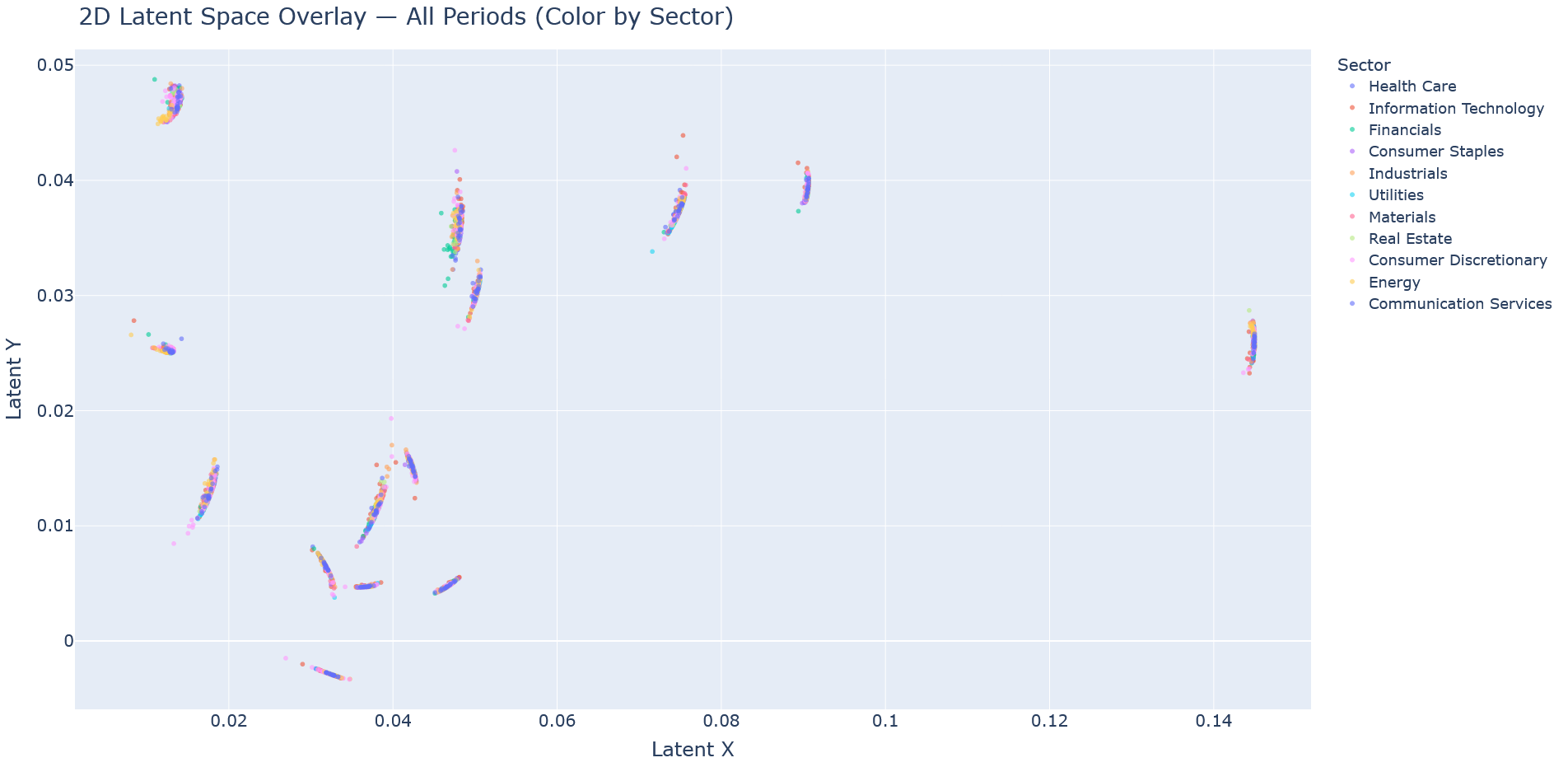}
    \caption{Two-dimensional projection of latent embeddings across all rolling periods.  
    Each point represents a stock in the latent space; color intensity indicates RBF kernel density.  
    Dense regions correspond to highly correlated temporal behaviors.}
    \label{fig:overlay_2d_scatter_all_periods}
\end{figure*}

\begin{figure*}[!b]
    \centering
    \includegraphics[width=\textwidth]{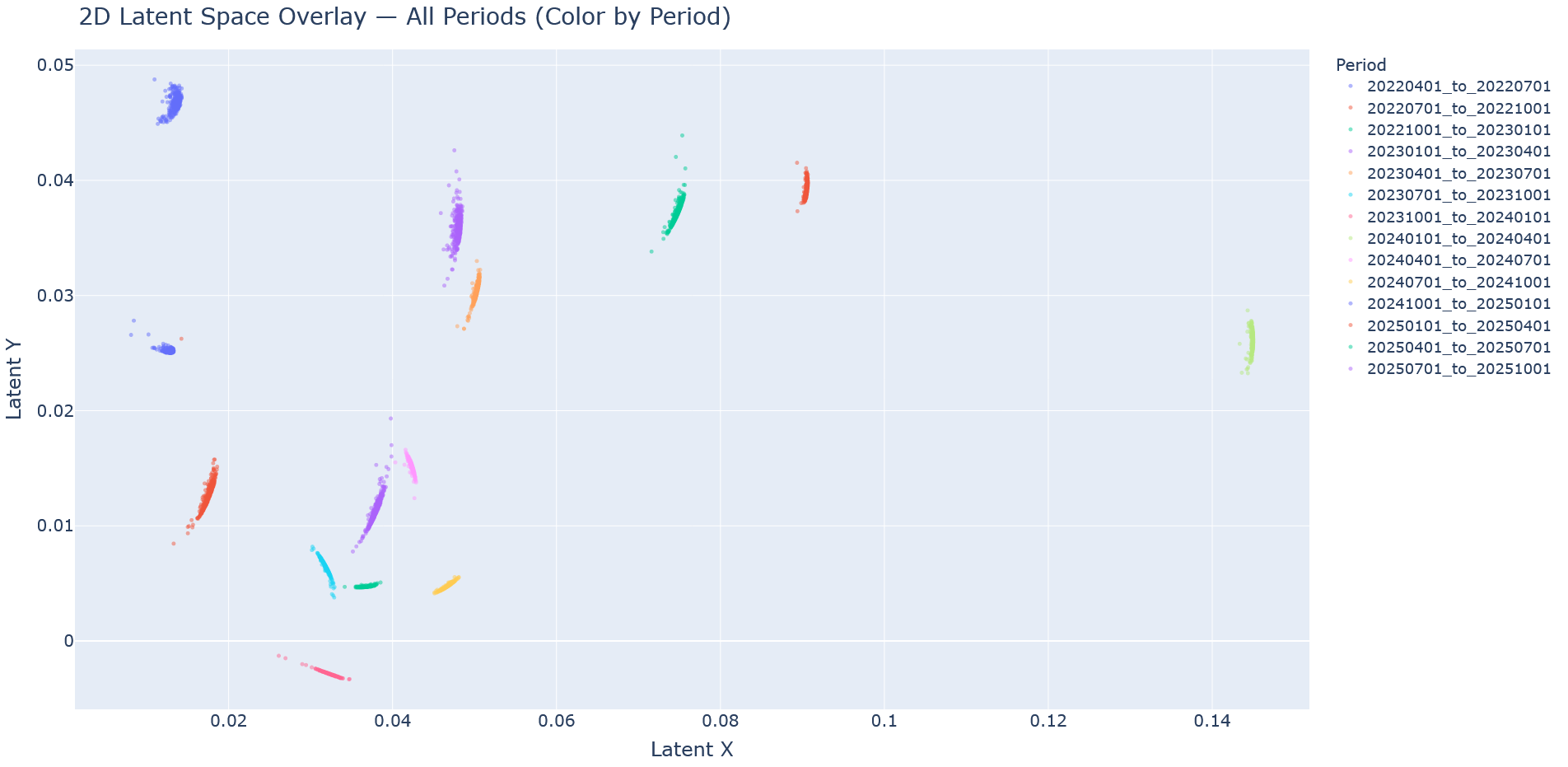}
    \caption{Latent embeddings for each quarterly test window.  
    Colors correspond to industry sectors.  
    The stable geometric structure across periods confirms the temporal generalization of QLSTM representations.}
    \label{fig:overlay_2d_scatter_all_periods_by_period}
\end{figure*}

\begin{figure*}[!b]
    \centering
    \includegraphics[width=\textwidth]{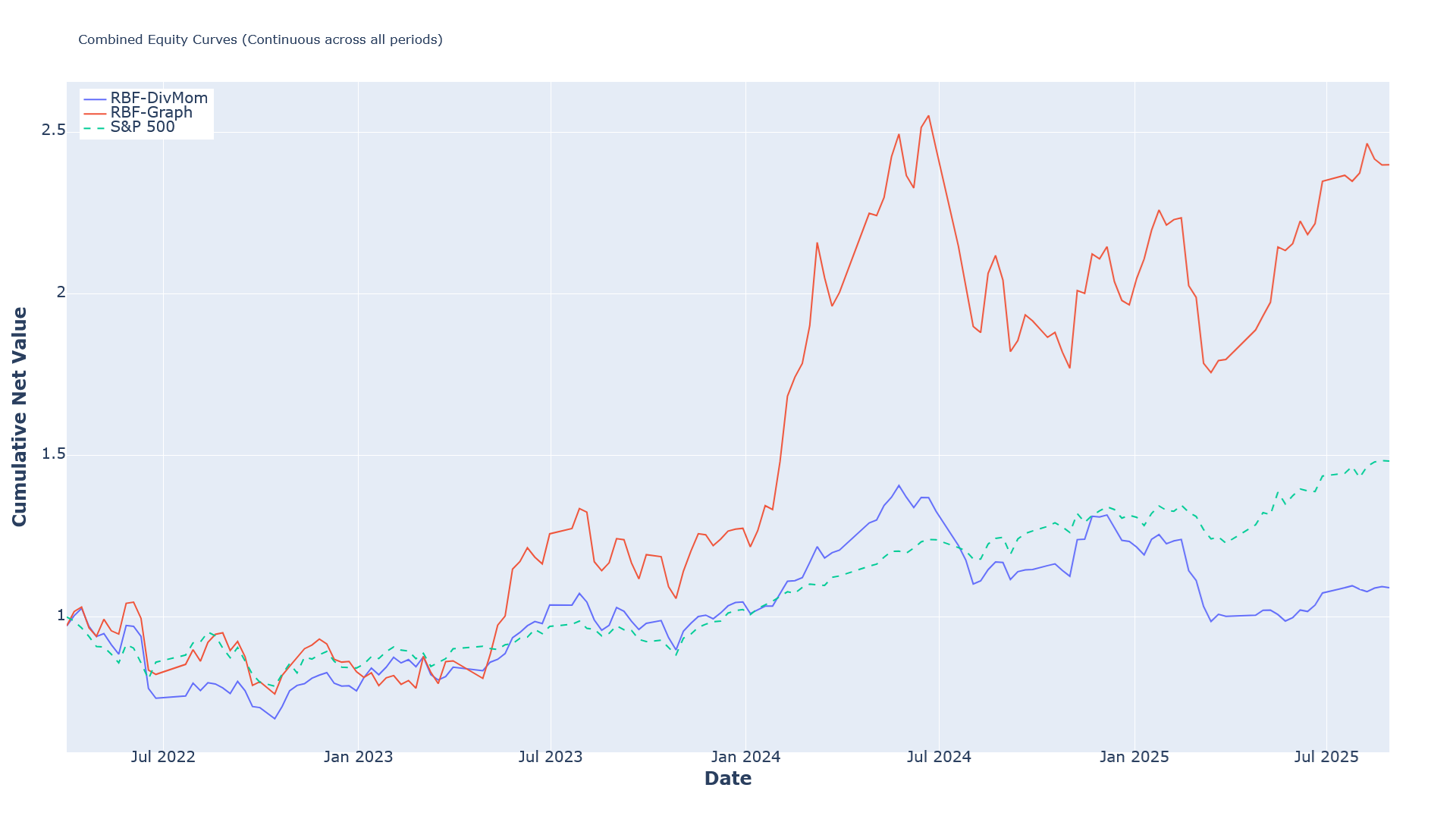}
    \caption{Cumulative equity curves of RBF-DivMom, RBF-Graph, and the S\&P~500 benchmark from 2022 to 2025.  
    The QLSTM–RBF models are retrained each quarter and compounded across test periods.  
    RBF-Graph achieves the highest cumulative return, while RBF-DivMom exhibits smoother growth and smaller drawdowns.}
    \label{fig:combined_equity_curves_largefont}
\end{figure*}

This section reports the empirical results of the proposed QLSTM–RBF framework.  
All experiments use weekly S\&P~500 constituent returns from 2022 to 2025 and are evaluated with rolling quarterly backtests to ensure temporal robustness.

\subsection{Latent Space Visualization and Interpretability}

In this latent plane, stocks form clear sector-based clusters, and Table~\ref{tab:portfolio_characteristics} summarizes how dominant clusters, sector composition, and market context relate to the behavior of the RBF-DivMom and RBF-Graph portfolios across rolling windows.

In Fig.~\ref{fig:overlay_2d_scatter_all_periods}, regions with high RBF density reveal groups of stocks sharing highly correlated return dynamics.  
Fig.~\ref{fig:overlay_2d_scatter_all_periods_by_period} further shows that this latent topology remains stable across quarters, indicating that the QLSTM encoder captures market structure that is robust to short-term noise and regime changes.

\subsection{Temporal Consistency Across Rolling Windows}

The quarterly rolling design yields 13 consecutive train–test pairs from 2022Q1 to 2025Q3.  
Despite macroeconomic regime shifts (e.g., inflation cycles and policy tightening), the latent RBF clusters preserve a consistent geometric layout over time.  
This temporal consistency allows portfolios constructed from the latent manifold to be compared across periods without additional alignment steps and supports the use of a single kernel-based diversification logic throughout the backtest horizon.

\subsection{Portfolio Performance Validation}

Two portfolio strategies are evaluated against the S\&P~500 benchmark:  
(1) \textit{RBF-DivMom}, a discrete momentum selector regularized by kernel similarity, and  
(2) \textit{RBF-Graph}, a continuous, graph-regularized optimizer using the same RBF kernel.  
Fig.~\ref{fig:combined_equity_curves_largefont} shows that both strategies outperform the benchmark, with RBF-Graph achieving the highest terminal equity and RBF-DivMom delivering smoother compounding and smaller drawdowns.

Table~\ref{tab:metrics_full} reports per-quarter excess statistics (CAGR, volatility, Sharpe ratio, and maximum drawdown) for RBF-DivMom, RBF-Graph, and the S\&P~500.  
Absolute performance is reconstructed by compounding each quarter’s output, leading to final net values of approximately 1.10$\times$ (DivMom), 2.40$\times$ (Graph), and 1.45$\times$ (S\&P) over the 2022–2025 period.  
These results indicate that the latent RBF kernel translates temporal representations into allocation signals that are both profitable and interpretable.

\begin{table*}[t]
\centering
\footnotesize
\caption{Quarterly performance of QLSTM–RBF strategies compared with S\&P~500 benchmark (2022–2025)}
\label{tab:metrics_full}
\setlength{\tabcolsep}{1pt}
\renewcommand{\arraystretch}{1.1}
\resizebox{\textwidth}{!}{%
\begin{tabular}{l|*{12}{>{\centering\arraybackslash}p{1.25cm}}}
\toprule
\multirow{2}{*}{\textbf{Period}} &
\multicolumn{4}{c}{\textbf{RBF-DivMom (excess)}} &
\multicolumn{4}{c}{\textbf{RBF-Graph (excess)}} &
\multicolumn{4}{c}{\textbf{S\&P~500 (excess)}} \\
\cmidrule(lr){2-5} \cmidrule(lr){6-9} \cmidrule(lr){10-13}
 & CAGR & Vol & Sharpe & MaxDD & CAGR & Vol & Sharpe & MaxDD & CAGR & Vol & Sharpe & MaxDD \\
\midrule
2022Q2 & -0.6874 & 0.4444 & -2.3719 & -0.2720 & -0.5452 & 0.4572 & -1.4911 & -0.2140 & -0.4556 & 0.2713 & -2.1045 & -0.1933 \\
2022Q3 & -0.1927 & 0.2613 & -0.6985 & -0.1029 & -0.1462 & 0.3843 & -0.2321 & -0.1726 & -0.3162 & 0.2563 & -1.3608 & -0.1656 \\
2022Q4 &  0.2994 & 0.2440 &  1.1869 & -0.0709 &  0.1320 & 0.2746 &  0.5775 & -0.1116 &  0.2118 & 0.2212 &  0.9704 & -0.0600 \\
2023Q1 &  0.3743 & 0.2502 &  1.3912 & -0.0851 &  0.1206 & 0.3854 &  0.4674 & -0.0978 &  0.2556 & 0.1656 &  1.4549 & -0.0706 \\
2023Q2 &  0.1497 & 0.2039 &  0.7398 & -0.0608 &  0.3972 & 0.3147 &  1.2620 & -0.0910 &  0.1868 & 0.1544 &  1.2014 & -0.0529 \\
2023Q3 &  0.2842 & 0.2471 &  1.0789 & -0.0893 &  0.4178 & 0.3090 &  1.3521 & -0.1043 &  0.2586 & 0.1951 &  1.3257 & -0.0771 \\
2023Q4 &  0.1094 & 0.2362 &  0.3973 & -0.0719 &  0.2655 & 0.2992 &  0.8877 & -0.0837 &  0.1533 & 0.1733 &  0.9281 & -0.0599 \\
2024Q1 &  0.2529 & 0.2131 &  1.0745 & -0.0838 &  0.3732 & 0.2915 &  1.2009 & -0.1011 &  0.1659 & 0.1451 &  1.1274 & -0.0533 \\
2024Q2 &  0.1781 & 0.2332 &  0.7639 & -0.0894 &  0.2826 & 0.3080 &  0.9420 & -0.0999 &  0.1320 & 0.1581 &  0.8347 & -0.0612 \\
2024Q3 &  0.1420 & 0.2457 &  0.5785 & -0.0972 &  0.2413 & 0.3228 &  0.8284 & -0.1083 &  0.0959 & 0.1694 &  0.6551 & -0.0688 \\
2024Q4 &  0.2247 & 0.2345 &  0.9578 & -0.0841 &  0.3311 & 0.3011 &  1.1003 & -0.0997 &  0.1531 & 0.1576 &  0.9248 & -0.0595 \\
2025Q1 &  0.1755 & 0.2264 &  0.7753 & -0.0883 &  0.3072 & 0.2969 &  0.9730 & -0.1032 &  0.1194 & 0.1460 &  0.7825 & -0.0560 \\
2025Q2 &  0.1869 & 0.2307 &  0.8104 & -0.0905 &  0.2988 & 0.3025 &  1.0241 & -0.1061 &  0.1277 & 0.1493 &  0.8568 & -0.0578 \\
\midrule
\textbf{Mean (per-quarter)} & 0.1750 & 0.2291 & 0.6965 & -0.0907 & \textbf{0.9845} & 0.3609 & 1.0703 & -0.1143 & 0.1833 & 0.1548 & 1.3152 & -0.0610 \\
\midrule
\multicolumn{13}{c}{\textbf{Final Net Value (from Fig.~\ref{fig:combined_equity_curves_largefont}):} \textbf{DivMom = 1.10×, Graph = 2.40×, S\&P = 1.45×}} \\
\bottomrule
\end{tabular}%
}
\vspace{1mm}
\begin{flushleft}
\footnotesize{\textit{Note.} Table reports \emph{per-window} (quarterly) averages of excess-return statistics. 
The final net values are \emph{compounded} portfolio levels from the same dataset used to plot Fig.~\ref{fig:combined_equity_curves_largefont}; 
therefore mean-based metrics and cumulative curves may differ.}
\end{flushleft}
\end{table*}

\begin{table*}[!b]
\centering
\caption{Quarterly qualitative characteristics aligned with observed performance (2022Q2--2025Q2)}
\label{tab:portfolio_characteristics}
\renewcommand{\arraystretch}{1.15}
\setlength{\tabcolsep}{6pt}
\scriptsize
\begin{tabularx}{\textwidth}{%
  >{\raggedright\arraybackslash}p{1.1cm}
  >{\raggedright\arraybackslash}p{1.6cm}
  >{\raggedright\arraybackslash}p{1.8cm}
  >{\centering\arraybackslash}p{1.0cm}
  >{\raggedright\arraybackslash}X}
\toprule
\textbf{Quarter} & \textbf{Market Regime} & \textbf{Main Sectors} & \textbf{Lead} & \textbf{Performance Interpretation} \\
\midrule
2022Q2 & Volatile recovery & Energy, Utilities & DivMom &
Highly dispersed latent clusters under inflation shock; DivMom's similarity penalty avoided overloading correlated names and slightly cushioned the broad market drawdown. \\\midrule
2022Q3 & Bearish phase & Energy, Staples & Graph &
Energy\mbox{-}centred clusters remained coherent on the manifold; Graph concentrated into these resilient regions and achieved smaller drawdowns than DivMom and the benchmark. \\\midrule
2022Q4 & Early rebound & IT, Discretionary & Graph &
Growth sectors formed a tight high\mbox{-}momentum cluster; Graph leaned into this compressed region while controlling redundancy, capturing most of the early rebound. \\\midrule
2023Q1 & Growth rally & IT, Comm. & Graph &
A strong growth rally produced very coherent IT and communication clusters; Graph's concentration on this dominant manifold region delivered the highest risk-adjusted returns. \\\midrule
2023Q2 & Sideways market & Industrials, Financials & DivMom &
Cluster boundaries became fuzzy in a sideways regime; DivMom spread risk across weakly structured regions and avoided committing to any unstable theme. \\\midrule
2023Q3 & Rate plateau & Financials, Energy & Graph &
Value sectors formed stable latent centroids with moderate momentum; Graph tracked these centroids more aggressively and extracted consistent excess return. \\\midrule
2023Q4 & Mixed signals & Healthcare, Staples & DivMom &
Manifold dispersion increased across defensive sectors; DivMom's stronger diversification reduced correlation exposure and produced smoother equity curves. \\\midrule
2024Q1 & Bullish trend & IT, Comm. & Graph &
Bullish conditions re\mbox{-}compressed growth clusters; Graph focused weight inside the most coherent high\mbox{-}momentum region and outperformed both DivMom and the index. \\\midrule
2024Q2 & Rotation to value & Industrials, Energy & DivMom &
Cluster centres drifted from growth to cyclicals; DivMom maintained balanced exposure across several emerging clusters and sidestepped sharp style\mbox{-}rotation noise. \\\midrule
2024Q3 & Consolidation & Materials, Staples & DivMom &
Trend signals weakened and clusters flattened; DivMom prioritised low\mbox{-}similarity selections, which improved drawdown control in a low\mbox{-}alpha environment. \\\midrule
2024Q4 & Pre\mbox{-}easing rally & IT, Discretionary & Graph &
Pre\mbox{-}easing optimism tightened growth clusters again; Graph increased exposure to these persistent momentum regions and extended its cumulative advantage. \\\midrule
2025Q1 & Volatility return & Energy, Utilities & DivMom &
Renewed volatility widened the manifold and blurred sector boundaries; DivMom cushioned losses by avoiding heavy concentration in any single volatile cluster. \\\midrule
2025Q2 & Late\mbox{-}cycle strength & Industrials, IT & Graph &
Late\mbox{-}cycle strength produced renewed, moderately tight clusters in industrials and IT; Graph exploited this structure to further widen its lead over the benchmark. \\
\bottomrule
\end{tabularx}
\end{table*}

\noindent
Overall, \textbf{RBF-Graph} delivers the strongest growth in cumulative wealth, while \textbf{RBF-DivMom} offers more conservative yet still superior performance relative to buy-and-hold.  
The combination of a stable latent manifold and an RBF-based diversification penalty appears to be crucial for achieving both robustness and interpretability.

\subsection{Relationship Between RBF-DivMom and RBF-Graph}
\begin{table}[ht]
\centering
\footnotesize
\caption{Grid search of RBF-DivMom penalty $\lambda$ compared with RBF-Graph}
\label{tab:grid_search_divmom_simple}
\setlength{\tabcolsep}{5pt}
\renewcommand{\arraystretch}{1.15}
\begin{tabular}{c|c|c|c}
\toprule
\textbf{DivMom $\lambda$} & 
\textbf{Mean Sharpe} &
\textbf{Gap to Graph} &
\textbf{Final Net Value} \\
\midrule
0.00 & 0.96 & 0.11 & 1.41$\times$ \\
0.15 & 0.95 & 0.12 & 1.36$\times$ \\
0.30 & 0.91 & 0.16 & 1.40$\times$ \\
0.45 & 0.86 & 0.21 & 1.33$\times$ \\
0.60 & 0.87 & 0.20 & 1.24$\times$ \\
0.75 & 0.73 & 0.34 & 1.09$\times$ \\
0.90 & 0.68 & 0.39 & 1.15$\times$ \\
1.00 & 0.65 & 0.42 & 0.98$\times$ \\
\midrule
\textbf{RBF-Graph} & \textbf{1.07} & \textbf{---} & \textbf{2.40$\times$} \\
\bottomrule
\end{tabular}
\end{table}
To assess how closely the discrete RBF-DivMom strategy approximates the continuous RBF-Graph solution, we conduct a grid search over $\lambda \in {0.00,0.15,\dots,1.00}$ across all fourteen rolling windows. Minor penalties ($\lambda \le 0.30$) yield the highest mean Sharpe ratios and the smallest gap to RBF-Graph, while larger penalties markedly reduce performance.
We therefore adopt $\lambda = 0.15$ as the representative DivMom setting. The behavior of DivMom reflects a clear trade-off: tiny $\lambda$ leads to momentum-heavy allocations with limited diversification, whereas large $\lambda$ over-diversifies and weakens returns. Near $\lambda=0.15$, the strategy best balances these effects, producing equity curves that closely follow the turning points of RBF-Graph and showing the smallest Sharpe discrepancy. The main distinction is adaptivity. RBF-Graph implicitly adjusts its effective regularization through continuous optimization of the kernel matrix, whereas DivMom relies on a fixed $\lambda$ across all periods. Thus, DivMom can approximate but not replicate Graph’s dynamic response to evolving latent geometry. The grid search confirms (see Table~\ref{tab:grid_search_divmom_simple}) that both strategies stem from the same RBF-driven diversification principle, differing primarily in flexibility rather than intent.

\section{Discussion}
Fig.~\ref{fig:overlay_2d_scatter_all_periods} and Fig.~\ref{fig:overlay_2d_scatter_all_periods_by_period} show that the quantum-enhanced encoder yields smooth trajectories and sector-coherent clusters across all fourteen rolling windows, remaining stable even in volatile quarters (e.g., 2022Q3, 2023Q2, 2024Q3) where the classical LSTM becomes less informative. This suggests that embedding a depth-1 variational circuit within each recurrent gate provides an additional bounded nonlinearity that acts as a compact feature expander under NISQ constraints and an implicit regularizer against noise-driven distortions in the latent space. Consistently, Fig.~\ref{fig:combined_equity_curves_largefont} and Table~\ref{tab:metrics_full} show that both kernel-based strategies benefit from these embeddings: Graph achieves the strongest long-run growth, while DivMom improves compounding smoothness and drawdown behaviour.

The latent geometry also admits an economic interpretation. Table~\ref{tab:portfolio_characteristics} indicates that when embeddings compress into tight clusters, Graph performs better because coherent similarity neighborhoods support concentrated allocation within stable temporal groups. When embeddings disperse, DivMom performs better because dispersion reflects fragmented dynamics (e.g., sector rotation or correlation breakdown) and benefits from diversification control. In this view, manifold compression corresponds to regimes dominated by a stronger common market mode (higher co-movement), whereas dispersion corresponds to weaker or shifting leadership. DivMom's similarity penalty mitigates redundant exposure and crowding, linking geometry to economically meaningful allocation behaviour without claiming quantum advantage. Table~\ref{tab:grid_search_divmom_simple} further supports this connection: moderate penalties reduce the performance gap to Graph, consistent with the geometry patterns observed in Fig.~\ref{fig:overlay_2d_scatter_all_periods}.

Finally, the approach remains practical under real-world constraints. The hybrid encoder follows the same training pipeline as the classical baseline, with a constant-factor overhead from evaluating a fixed-depth (depth-1) quantum layer per time step. Keeping circuit depth shallow preserves feasibility for periodic retraining and time-sensitive rebalancing, suggesting a realistic role for hybrid quantum architectures in NISQ-era sequence modelling when data are limited and representations must remain stable for downstream decisions.

\section{Conclusion}

We show that hybrid quantum-classical sequence modelling can improve equity allocation under practical NISQ constraints. Across fourteen rolling quarters, the QLSTM Seq2Seq encoder learns latent manifolds that are more geometrically stable than a classical LSTM, with smoother trajectories, stronger sector coherence, and clearer regime separation. Because these embeddings define the RBF similarity kernel used for allocation, representation quality directly shapes portfolio decisions. Empirically, Tables~\ref{tab:metrics_full} and \ref{tab:portfolio_characteristics} show that quantum-enhanced embeddings provide a more reliable similarity basis than raw returns or classical embeddings, yielding stronger risk-adjusted performance across quarters. The results also explain when each optimiser works best: RBF-Graph dominates when the latent space compresses into tight, coherent clusters that support concentrated bets, while RBF-DivMom is more robust when embeddings disperse and cross-sector dynamics fragment, benefiting from diversification. Shallow variational circuits within recurrent gates efficiently and interpretably shape temporal geometry, improving allocation performance without claiming quantum advantage.

While we focus on a classical LSTM baseline to isolate the effect of the shallow quantum augmentation, future work will benchmark against stronger time-series representation models (e.g., transformer-based architectures) and self-supervised embedding methods. We will also report additional finance metrics (e.g., Sortino ratio and tracking error) and evaluate scalability on broader universes (ETFs/crypto) and longer horizons. Finally, we will explore circuit ansatz/encoding choices tailored to financial data distributions to further understand when and why shallow VQCs act as effective regularizers.

\bibliographystyle{IEEEtran}
\bibliography{references}

\begin{thebibliography}{10}
\providecommand{\url}[1]{#1}
\csname url@samestyle\endcsname
\providecommand{\newblock}{\relax}
\providecommand{\bibinfo}[2]{#2}
\providecommand{\BIBentrySTDinterwordspacing}{\spaceskip=0pt\relax}
\providecommand{\BIBentryALTinterwordstretchfactor}{4}
\providecommand{\BIBentryALTinterwordspacing}{\spaceskip=\fontdimen2\font plus
\BIBentryALTinterwordstretchfactor\fontdimen3\font minus \fontdimen4\font\relax}
\providecommand{\BIBforeignlanguage}[2]{{%
\expandafter\ifx\csname l@#1\endcsname\relax
\typeout{** WARNING: IEEEtran.bst: No hyphenation pattern has been}%
\typeout{** loaded for the language `#1'. Using the pattern for}%
\typeout{** the default language instead.}%
\else
\language=\csname l@#1\endcsname
\fi
#2}}
\providecommand{\BIBdecl}{\relax}
\BIBdecl

\bibitem{9747369}
S.~Y.-C. Chen, S.~Yoo, and Y.-L.~L. Fang, ``Quantum long short-term memory,'' in \emph{ICASSP 2022 - 2022 IEEE International Conference on Acoustics, Speech and Signal Processing (ICASSP)}, 2022, pp. 8622--8626.

\bibitem{e26110954}
\BIBentryALTinterwordspacing
K.~Kea, D.~Kim, C.~Huot, T.-K. Kim, and Y.~Han, ``A hybrid quantum-classical model for stock price prediction using quantum-enhanced long short-term memory,'' \emph{Entropy}, vol.~26, no.~11, 2024. [Online]. Available: \url{https://www.mdpi.com/1099-4300/26/11/954}
\BIBentrySTDinterwordspacing

\bibitem{mironowicz2024applicationsquantummachinelearning}
\BIBentryALTinterwordspacing
P.~Mironowicz, A.~S. H., A.~Mandarino, A.~E. Yilmaz, and T.~Ankenbrand, ``Applications of quantum machine learning for quantitative finance,'' 2024. [Online]. Available: \url{https://arxiv.org/abs/2405.10119}
\BIBentrySTDinterwordspacing

\bibitem{s24247932}
\BIBentryALTinterwordspacing
J.~Lee, I.~Ham, Y.~Kim, and H.~Ko, ``Time-series representation feature refinement with a learnable masking augmentation framework in contrastive learning,'' \emph{Sensors}, vol.~24, no.~24, 2024. [Online]. Available: \url{https://www.mdpi.com/1424-8220/24/24/7932}
\BIBentrySTDinterwordspacing

\bibitem{Kong2025}
\BIBentryALTinterwordspacing
X.~Kong, Z.~Chen, W.~Liu, K.~Ning, L.~Zhang, S.~Muhammad~Marier, Y.~Liu, Y.~Chen, and F.~Xia, ``Deep learning for time series forecasting: a survey,'' \emph{International Journal of Machine Learning and Cybernetics}, vol.~16, no. 7--8, pp. 5079--5112, Feb. 2025. [Online]. Available: \url{http://dx.doi.org/10.1007/s13042-025-02560-w}
\BIBentrySTDinterwordspacing

\bibitem{zaman2024poqaframeworkportfoliooptimization}
\BIBentryALTinterwordspacing
K.~Zaman, A.~Marchisio, M.~Kashif, and M.~Shafique, ``Po-qa: A framework for portfolio optimization using quantum algorithms,'' 2024. [Online]. Available: \url{https://arxiv.org/abs/2407.19857}
\BIBentrySTDinterwordspacing

\bibitem{lu2024quantuminspiredportfoliooptimizationqubo}
\BIBentryALTinterwordspacing
Y.-C. Lu, C.-M. Fu, L.-P. Yu, Y.-J. Chang, and C.-R. Chang, ``Quantum-inspired portfolio optimization in the qubo framework,'' 2024. [Online]. Available: \url{https://arxiv.org/abs/2410.05932}
\BIBentrySTDinterwordspacing

\bibitem{Sakuler2025}
\BIBentryALTinterwordspacing
W.~Sakuler, J.~M. Oberreuter, R.~Aiolfi, L.~Asproni, B.~Roman, and J.~Schiefer, ``A real-world test of portfolio optimization with quantum annealing,'' \emph{Quantum Machine Intelligence}, vol.~7, no.~1, p.~43, 2025. [Online]. Available: \url{https://doi.org/10.1007/s42484-025-00268-2}
\BIBentrySTDinterwordspacing

\bibitem{Naik2025}
\BIBentryALTinterwordspacing
A.~S. Naik, E.~Yeniaras, G.~Hellstern, G.~Prasad, and S.~K. L.~P. Vishwakarma, ``From portfolio optimization to quantum blockchain and security: a systematic review of quantum computing in finance,'' \emph{Financial Innovation}, vol.~11, no.~1, p.~88, 2025. [Online]. Available: \url{https://doi.org/10.1186/s40854-025-00751-6}
\BIBentrySTDinterwordspacing

\bibitem{Mengara2025}
O.~Mengara, ``The art of quantum computing for finance: Brief overview and prospects,'' \emph{SSRN Electronic Journal}, Jul. 2025, available at SSRN: \url{https://ssrn.com/abstract=5373357} or \url{http://dx.doi.org/10.2139/ssrn.5373357}.

\end{thebibliography}

\end{document}